\documentstyle[11pt]{article}

\newcommand{\eq}[1]{equation~(\ref{#1})}
\newcommand{\eqs}[2]{equations~(\ref{#1}) and~(\ref{#2})}
\newcommand{\eqto}[2]{equations~(\ref{#1}) to~(\ref{#2})}
\newcommand{\Ei}[1]{\mbox{Ei}{#1}}
\newcommand{\erf}[1]{\mbox{erf}{#1}}

\begin{document}

\thispagestyle{empty}

\mbox{}

{\raggedleft

WU-AP/53/95 \\
astro-ph/9511143 \\}

\vspace{7mm}

\begin{center}  {\Large\bf\expandafter{
       An Inflationary Model with an \\
       Exact Perturbation Spectrum }}
\end{center}

\vspace{10mm}

\begin{center}
{\Large Richard Easther} \footnote{ easther@cfi.waseda.ac.jp}\\
\medskip
Department of Physics, \\
Waseda University,  3-4-1 Okubo,  Shinjuku-ku, \\ Tokyo,  Japan.
\end{center}

\vspace{2cm}

\setcounter{footnote}{0}
\setcounter{page}{0}

\section*{Abstract}

We present a new, exact scalar field cosmology for which the spectrum
of scalar (density) perturbations can be calculated exactly. We use
this exact result to the probe the accuracy of approximate calculations
of the perturbation spectrum.

\vfill
\noindent PACS:  04.20.Jb  98.80.C

%

\mbox{}

\newpage

\section{Introduction}

The inflationary paradigm was originally motivated by its ability to
solve the ``initial conditions'' problems associated with the standard
model of the Big Bang~\cite{Guth1981a}. However it was quickly
realised that an inflationary epoch would also produce primordial
density perturbations and may be able to explain both the observed
clustering of galaxies and the (then unobserved) anisotropies in the
Cosmic Microwave Background (CMB).

Insisting that inflation produces the observed spectrum of
primordial perturbations is a more demanding requirement than
merely providing the approximately 60 e-foldings of inflation needed
to solve the various initial conditions problems.  Consequently, the
focus of much present work is on the density perturbation spectra
produced by different inflationary models.  This is particularly true
of slow-rolling inflation, in which the scalar field evolves
continuously.  The consistency of slow-rolling inflation can be
directly tested through CMB observations \cite{DodelsonET1994a}, and
in principle the potential can be reconstructed
\cite{CopelandET1993a,CopelandET1993c}, opening a window into a
GUT-scale particle physics.

In order to do this, accurate calculations of the perturbation spectra
produced during inflation are required.  Stewart and Lyth
\cite{StewartET1993a} give a second order calculation of the
perturbation spectra for a general potential.  Exact scalar field
cosmologies have been widely studied, in for instance,
\cite{Madsen1986a,Muslimov1990a,EllisET1991a,Lidsey1991a,Easther1993b,%
Barrow1993a,Barrow1994a}, but power-law inflation
\cite{LucchinET1985a,LucchinET1985b,SalopekET1990a} remains the only
only inflationary model for which the perturbation spectrum has been
obtained exactly \cite{AbbottET1984a,LythET1992a}.

The purpose of this paper is to present a new scalar field cosmology
for which the spectrum of scalar perturbations can be obtained
analytically. This solution therefore provides a second example of a
slow-rolling inflationary cosmology with an exact perturbation
spectrum. While the perturbations produced are not within the
parameter range permitted by observation, this model extends our
understanding of the problem and can be used to probe the validity of
the approximation schemes used to tackle the more general problem.

\section{The Equations of Motion}

For a scalar field, $\phi$ with potential $V(\phi)$ in a spatially
flat Robertson-Walker metric we have
\begin{eqnarray}
H^2  &=& \frac{1}{3}\left(\frac{\dot{\phi}^2}{2} + V(\phi)\right),
   \label{Hsqrd} \\
\frac{\ddot{a}}{a} &=& \frac{1}{3} \left(V(\phi) - \dot{\phi}^2\right),
    \label{adda}
\end{eqnarray}
where $a$ is the scale factor and $H = d \ln{a} / dt$, is the Hubble
parameter. {}From these equations we obtain the equation of motion
for the scalar field,
\begin{equation}
\ddot{\phi} + 3H\dot{\phi} + \frac{dV(\phi)}{d\phi} = 0. \label{eofmotion}
\end{equation}

As is often the case when dealing with exact scalar field cosmologies
it will be useful to parametrise the motion in terms of the field,
$\phi$ \cite{Muslimov1990a,Easther1993b,SalopekET1990a,Lidsey1991b}.
{}From \eqs{Hsqrd}{adda} we deduce that $d\phi /dt = -2d H/d\phi$,
leading to
\begin{eqnarray}
V(\phi) &=& 3H^2 - 2H'^{2},
       \label{Vflatphi} \\
\frac{a}{a_0} &=&
    \exp{\left(-\frac{1}{2}\int_{\phi_0}^\phi{\frac{H}{H'}d\phi}\right)},
          \label{aflatphi} \\
t &=& -\frac{1}{2}\int_{\phi_0}^\phi{\frac{1}{H'}d\phi},
    \label{tflatphi}
\end{eqnarray}
where a dash denotes differentiation with respect to $\phi$.  If we
specify $H(\phi)$ we can immediately obtain the corresponding
potential and evolution. The equation governing the evolution of
scalar perturbations with wavenumber $k$ is
\cite{StewartET1993a,Mukhanov1989a,MakinoET1991a,MukhanovET1992b}
\begin{equation}
\frac{d^2 u_k}{d \eta^2} +
   \left( k^2 - \frac{1}{z}\frac{d^2 z}{d \eta^2} \right) u_k = 0
  \label{mode1}
\end{equation}
where $\eta = \int{1/a dt}$ is the conformal time and
$z = a\dot{\phi}/H$. Furthermore, we have the boundary conditions
\begin{eqnarray}
u_k \to \frac{1}{\sqrt{2k}}e^{-i k \eta} \ &,& \ aH \ll k,
   \label{bcond1} \\
u_k \propto z \ &,& \ aH \gg k,
    \label{bcond2}
\end{eqnarray}
which guarantees that the perturbation behaves like a free field well
inside the horizon and is fixed at superhorizon scales.

In practice, however, we are interested in the spectrum $P_s$
and index, $n_s$ which are given by
\begin{eqnarray}
P_{s}^{1/2}(k) &=& \left. \sqrt{\frac{k^3}{2\pi^2}}
     \left|\frac{u_k}{z}\right| \right|_{aH = k}, \\
n_{s} &=& 1 + \frac{d \ln P_{s}}{d\ln{k}} .
\end{eqnarray}
The form of \eq{mode1} can be simplified by defining $U_k = u_k /z$,
and
\begin{eqnarray}
&& \frac{d^2 U_k}{d \eta^2} +
   \frac{2}{z}\frac{dz}{d\eta} \frac{d U_k}{d \eta} +
   k^2 U_k = 0  \nonumber \\
& \Rightarrow&
  \frac{d^2 U_k}{d \eta^2} +
   4 H a \left[ \frac{1}{2} + \left(\frac{H'}{H}\right)^2
              - \frac{H''}{H}\right] \frac{d U_k}{d \eta} +
   k^2 U_k = 0.  \label{mode2}
\end{eqnarray}
We now turn our attention to the particular case where $z$ is a
constant, which is equivalent to demanding that the term in square
brackets in \eq{mode2} vanishes.

\section{The Inflationary Model}

In order to construct a model with an exact perturbation spectrum we
demand that $z$ is constant. This is equivalent to requiring that $H$
satisfy the differential equation
\begin{equation}
\frac{1}{2} + \left(\frac{H'}{H}\right)^2
              - \frac{H''}{H} = 0 \label{Hcond}
\end{equation}
which has the solution
\begin{equation}
H(\phi) = B \exp{\left( \frac{\phi^2}{4} + A\phi\right)}.
   \label{Hsoln}
\end{equation}
The values of the integration constants $A$ and $B$ are arbitrary, but
we can set $A=0$ without loss of generality, as it can be recovered by
making a linear shift of the field. {}From \eqto{Vflatphi}{tflatphi} we
derive the corresponding exact scalar field cosmology,
\begin{eqnarray}
V(\phi) &=& B^2 \left( 3 -\frac{\phi^2}{2}\right)
           \exp{\left( \frac{\phi^2}{2} \right)}, \label{Vexact}  \\
a(\phi) &=& \frac{\phi_0}{\phi}, \label{aexact}  \\
t(\phi) &=& \frac{1}{2B} \left[ \Ei{\left(-\frac{\phi_0^2}{4}\right)} -
  \Ei{\left(-\frac{\phi^2}{4} \right)}\right], \label{texact}
\end{eqnarray}
where $\Ei{}$ is the exponential integral function. The conformal time
is
\begin{equation}
\eta(\phi) = \frac{B\sqrt{\pi}}{\phi_0}
\left[  \erf{\left(-\frac{\phi_0}{2}\right)} -
              \erf{\left(-\frac{\phi}{2}   \right)}\right] .
  \label{etaexact}
\end{equation}
At late times, or as $\phi$ goes to zero, the conformal time tends to
a constant value.

The cosmological properties of this solution are quickly derived. The
potential, shown in Figure~1, is not bounded below. However, for
this solution the total energy density is always positive as the
kinetic energy is very large in the region where $|\phi|$ is large.
The motion is not inflationary at all times. By definition, inflation
occurs when $\ddot{a} > 0$, or when $\epsilon < 1$, where
\begin{equation}
\epsilon = 2\left(\frac{H'}{H}\right)^2. \label{epflatphi}
\end{equation}
Thus inflation occurs only when $|\phi| < \sqrt{2}$. If this model was
to produce {\em all\/} the 60 e-foldings of inflation needed to solve
the initial conditions problems in the standard model of cosmology,
$\phi$ must evolve to be unreasonably close to zero in view of the
typical size of the perturbations in the field.

This exact inflationary model is similar to one previously discussed
by Barrow \cite{Barrow1994a}, and can clearly be generalised in a
number of ways.  In particular, taking the Hubble parameter to be of
the form $H(\phi) = B\exp{(\phi^2/m)}$ gives a similar potential.
However, in this paper we will focus on the case where $z$ is
constant, which requires $H$ to have the form of \eq{Hsoln}.

\section{The Perturbation Spectrum}

We now turn our attention to the spectrum of scalar perturbations
produced by this model. The solution to \eq{mode2} is simple, as the
first derivative term drops out and we find
\begin{equation}
u_k(\eta) = \frac{1}{\sqrt{2k}} e^{-i k \eta} \label{Usoln}
\end{equation}
for the growing mode, after we have imposed the boundary conditions.
In terms of the conformal time, this solution has the special property
that the perturbations always evolve according to the equation of
motion for a free field. However, since $\eta$ tends to a constant
value $u_k$ is essentially fixed at super-horizon scales, as we would
expect.

We can immediately calculate the spectral index for these
perturbations, giving
\begin{equation}
n_s = 3.
\end{equation}
Note that this value is both exact and independent of  $\phi$. A flat,
or scale-free,  spectrum has an index of unity. Hence this spectrum is
``blue''  as it possesses more  power at large values of $k$, or small
scales.   Astrophysical   constraints,   both   {}from   the  CMB   data
\cite{BennettET1994a} and bounds on  the density  of small  primordial
black holes \cite{CarrET1994a} rule out such a large value of $n_{s}$
in the physical universe. The spectrum of the tensor, or
gravitational, perturbations has not been obtained exactly. However
when $\phi \approx 0$, the expansion is roughly exponential and we
therefore expect the tensor perturbation spectrum to be flat.

This model can be used to probe the accuracy of the first and second
order expressions for the scalar perturbation spectra. Since this
exact result pertains to a particular potential it does not provide a
definitive test, but the results are reassuring. Written in terms of
derivatives of $H(\phi)$ the perturbation spectrum
is~\cite{StewartET1993a}
\begin{eqnarray}
n_{s,1}(\phi) &=& 1 - 8\left( \frac{H'}{H}\right)^2 - 4\frac{H''}{H},
     \\
n_{s,2}(\phi) &=& n_{s,1} - 8(1+c)\left( \frac{H'}{H}\right)^4
   -2(3-5c)\frac{H'^2 H''}{H^3} + 2(3-c)\frac{H'''H'}{H^2}.
\end{eqnarray}
The subscripts 1 and 2 denote the first and second order values,
respectively, and $c = 4(\ln{2} + \gamma) - 5 \approx 0.081$, where
$\gamma$ is the Euler-Mascheroni constant.

In Figure~2 the first and second order results are plotted for
the values of $\phi$ during which this solution is inflationary.  We
can immediately see that the second order result gives a much more
accurate estimate of the index than the first order value, remaining
within approximately 10\% of the exact value when $\phi \leq 1$.  The
departures from slow roll for this potential are large, even
though the expansion of the scale factor is approximately exponential.

\section{Discussion}

The exact scalar field cosmology developed in this paper adds another
specimen to the collection of analytic inflationary models that have
been discovered over the last decade. Unusually, however, the mode
equation for the scalar perturbations can also be solved, making this
only the second model (after power-law inflation) for which the exact
value of the spectral index of the scalar perturbations is
known. The spectrum has an index of 3, which is significantly
different {}from the scale free value of unity, and is too large to
satisfy the observational bounds on the primordial spectrum. However,
this solution is useful as it can be used to test approximate
calculations of the spectral indices for slow-rolling inflation. Since
the slow-rolling conditions are badly violated by this model the test
is a comparatively stringent one, and the first order expression does
not give good agreement with the exact result. However, the second
order expression for the scalar index is able to match the exact
result to within 10\% over most of the inflationary epoch, a result
which will add confidence to its use in more realistic situations.

\section*{Acknowledgements}

The author thanks Ed Copeland for a valuable discussion and is
supported by a JSPS post-doctoral fellowship and the Grant-in-Aid for
JSPS fellows (0694194).

\section*{Figure Captions}

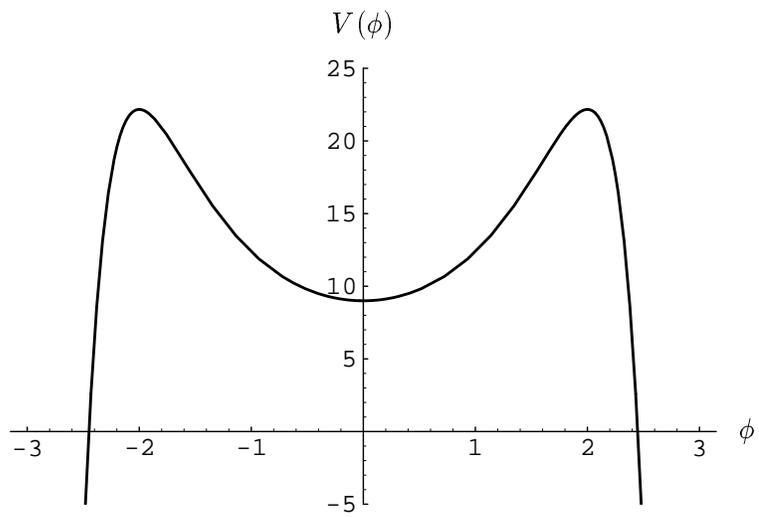
\noindent {\bf Figure 1}
The potential, \eq{Vexact}, is plotted for $B=1$. The
solution is only inflationary when $\phi$ is near the origin.

\vspace{2cm}

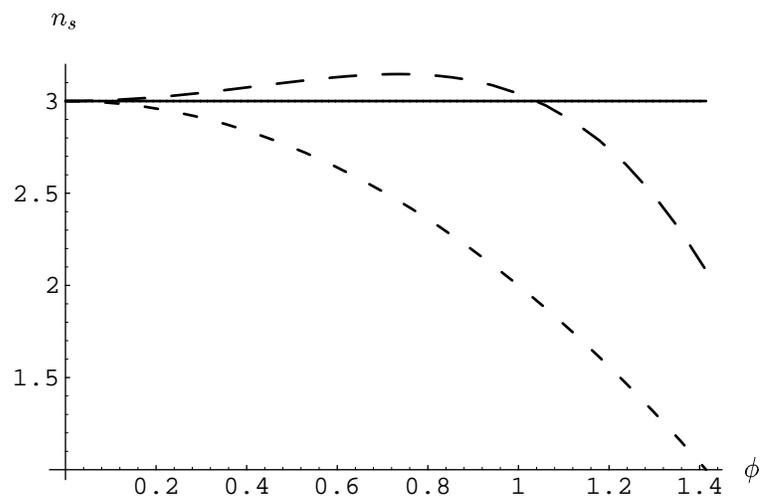
\noindent {\bf Figure 2}
The first order (short dashes) and second order
(long dashes) results for the perturbation spectrum are plotted, with
the exact value of 3 shown for reference, against the values of $\phi$
for which the exact solution is inflationary.

\end{document}